\documentclass[aps,pre,a4,floatfix,groupedaddress,showpacs,showkeys,twocolumn]{revtex4-1}  
\usepackage{graphicx,amsmath,amssymb,dsfont,bm}

\usepackage{color}

\begin{document}

\title{Price response in correlated financial markets: empirical results}
\author{Shanshan Wang}
\email{shanshan.wang@uni-due.de}
\affiliation{Fakult\"at f\"ur Physik, Universit\"at Duisburg--Essen, Lotharstra\ss e 1, 47048 Duisburg, Germany}
\author{Rudi Sch\"afer}
\affiliation{Fakult\"at f\"ur Physik, Universit\"at Duisburg--Essen, Lotharstra\ss e 1, 47048 Duisburg, Germany}
\author{Thomas Guhr}
\affiliation{Fakult\"at f\"ur Physik, Universit\"at Duisburg--Essen, Lotharstra\ss e 1, 47048 Duisburg, Germany}

\date{\today}

\begin{abstract}
Previous studies of the stock price response to individual trades
  focused on single stocks. We empirically investigate the price
  response of one stock to the trades of other stocks. How large is
  the impact of one stock on others and vice versa?  --- This impact
  of trades on the price change across stocks appears to be transient instead of
  permanent. Performing different averages, we distinguish active and
  passive responses. The two average responses show different
  characteristic dependences on the time lag. The passive response
  exhibits a shorter response period with sizeable volatilities, and
  the active response a longer period.  We also study the response for
  a given stock with respect to different sectors and to the whole
  market. Furthermore, we compare the self--response with the various cross--responses.
  The correlation of the trade signs is a short--memory
  process for a pair of stocks, but it turns into a long--memory
  process when averaged over different pairs of stocks.   
\end{abstract} 
\pacs{ 89.65.Gh 89.75.Fb 05.10.Gg}

\keywords{econophysics, complex systems, statistical analysis }

\maketitle

\section{Introduction} 
\label{section1}

The trading at stock exchanges is organized by the order book whose
main purpose is to provide the same information to all market
participants. Although often ignored in the model building, it has a
large impact on the price dynamics and thus on the stylized facts as
well as on the more specific features~\cite{Cont2001,Chordia2002,Bouchaud2003,Bouchaud2008,Chakraborti2011,Toth2011,Eisler2012,Schmitt2012,Schmitt2013}. The stock
price is determined via a continuous double auction~\cite{Farmer2004},
in which some traders submit market orders for immediate transactions
at the best available price, while other traders submit limit orders
which specify an acceptable price for the trade. The limit orders are
listed in the order book.  Most of them do not immediately lead to
trades. The buy limit orders are referred to as bids and the sell
limit orders as asks. The best ask and best bid prices are the
quotes. Market orders do not appear in the order book. When a market
order is executed, it can either keep the quote unchanged, raise the
best ask price in case of a buy market order or lower the best bid
price for a sell market order. The prices change persistently as they
are affected by the incoming market orders. To profit from the price
difference between ask and bid, traders provide the limit orders which
leads to an anti--persistence of prices. As a result of a detailed
balance between persistent and anti--persistent, \textit{i.e.},
between super-- and subdiffusive behavior, the price on an intraday
scale moves diffusively like a random walk~\cite{Bouchaud2004}.

In recent years, a high autocorrelation of the order flow was
empirically found~\cite{Bouchaud2004,Lillo2004,Lillo2005,Toth2015}.
The splitting of orders over longer times introduces the long memory
of the order flow~\cite{Lillo2005} with remarkable persistence. Buy
(sell) orders are often followed by more buy (sell)
orders. Furthermore, the relation between trades and price changes has
received considerable attention~\cite{Chordia2002,Bouchaud2008,Bouchaud2004,Hausman1992,Kempf1999,Dufour2000,Plerou2001,Rosenow2002,Bouchaud2006,Mike2008} The Efficient Market
Hypothesis (EMH)~\cite{Fama1970} states that all available information
is processed and encoded in the current price, which would rule out any
(statistical) arbitrage opportunities. While this is in conflict with
the very different time scales on which, first, relevant new
information arrives and, second, the prices change, the model of Zero
Intelligence Trading (ZIT)~\cite{Gode1993} simply assumes randomly
acting trader, but also arrives at a memory--less random walk.

Based on the EMH, there are two major approaches to explain the impact
of trades on the stock price change. The first approach put forward by
Lillo and Farmer (LF)~\cite{Lillo2004}, suggests that the price impact is
permanent, but fluctuates with order size. The impact is caused by an
asymmetry in liquidity which is induced by the trade. The response
exhibits a power--law relation between order size and price
change~\cite{Lillo2004,Lillo2003,Gabaix2003,Plerou2004}. 
In the second approach, Bouchaud, Gefen, Wyart and Potters
(BGPW)~\cite{Bouchaud2004} argue that the price impact is transient,
but fixed with order size. The fact that the impact decays with time
is a result of price mean reversion. Moreover, they identify the
relation between order size and price response as
logarithmic~\cite{Potters2003}. Gerig~\cite{Gerig2008} suggests that
the two approaches LF and BGPW are equivalent and can be related by
exchanging variables. He also argues that the impact comes from
the asymmetric liquidity rather than the price mean reversion.

There are numerous studies devoted to the price response, but they all
focus on one single stock. Here, we go beyond this and investigate the
role of correlations. We carry out a large--scale empirical study of
real--time trade data and find a non--vanishing price response across
different stocks. We shed light on the price impact from trades in
different stocks by discussing the efficiency of the financial market
and by analyzing how the stocks respond to the whole market and to
different economic sectors. We thereby present a first complete view
of the response in the market as a whole and identify several
structural characteristics.

The paper is organized as follows. In Sect.~\ref{section2}, we present
our data set of stocks and provide some basic definitions.  In
Sect.~\ref{section3}, we show the empirical results, which indicate the
existence of trade sign correlation and price response in different
stocks, we also estimate the response noise. In Sect.~\ref{section4},
we discuss the trade impact on the prices from the viewpoint of market
efficiency. We introduce and work out two types of average response in
Sect.~\ref{section5}, an active and a passive one.  In Sect.~\ref{section6}, we compare the self--response with the various cross--responses. We give our conclusions in Sect.~\ref{section7}.

\section{Data description and time convention}   
\label{section2}

In Sect.~\ref{sec21}, we present the data set that we use in our
analysis. We discuss the proper choice of time convention in
Sect.~\ref{sec22}.

\subsection{Data set}
\label{sec21}

Our study is based on the data from NASDAQ stock market in the year
2008. NASDAQ is a purely electronic stock exchange, whose Trades and
Quotes (TAQ) data set contains the time, price and volume. This
information is not only given for the trades with all successive
transactions, but also for the quotes with all successive best buy and
sell limit orders.  

To investigate the response across different stocks in Sect.~\ref{section3}, we select six
companies from three different economic sectors traded in the NASDAQ
stock market in 2008. The stocks we analyzed are listed in
Table~\ref{table1} together with their acronyms and the corresponding
economic sectors.
\begin{table} [b]
\begin{center}
\caption{Company information} 
\begin{tabular}{lll}
\hline
\hline
Company					&Symbol&Sector	\\
\hline
Apple Inc.           		&AAPL   &Information technology\\
Microsoft Corp.      	&MSFT   &Information technology\\
Goldman Sachs Group&GS     &Financials\\
JPMorgan Chase       &JPM    &Financials\\
Exxon Mobil Corp.    	&XOM    &Energy\\
Chevron Corp.        	&CVX     &Energy\\
\hline
\hline
 \label{table1}
\end{tabular}
\end{center}
\vspace*{-0.6cm} 
\end{table}

When studying the market response in Sect.~\ref{section4}, we select the first ten stocks
with the largest average market capitalization in each economic sector
of the S$\&$P 500 index in 2008, except for the telecommunications services
where only nine stocks were available in that year. We recall that the market
capitalization is the trade price multiplied with the traded volume,
and the average is performed over every trade during the year 2008.
The selected 99 stocks are listed in App.~\ref{appA}. The 99 stocks are also ranked according to strongest passive and active responses in Sect.~\ref{sec52}.  

For the average responses of an individual stock in Sect.~\ref{section5}, the stocks AAPL, GS, XOM are selected as examples. The necessary average are performed over the remaining 495 stocks in the S$\&$P 500 index or over the stocks in a given economic sector. Here, we neglect the self-response of the stocks.

We only consider the common trading days in which the trading of stocks $i$ and $j$ took place. This is so because the trades of one stock $j$ in one day would not impact the intraday price of another stock $j$ without any trade in that day, and \textit{vice versa}.

\subsection{Physical versus trading time}
\label{sec22}

While studies on the response in single stocks typically employ trading
time as time axis, this is not useful when studying the response
across different stocks, because each stock has its own trading time.
Hence, we have to use the real, physical time. We project the data set
to a discrete time axis.  The quote data and the trade data of each
stock are in two separate files with a time--stamp accuracy of one
second. However, more than one quote or trade may be recorded in the
same second. Due to the one--second accuracy of the time--stamps, it
is not possible to match each trade with the directly preceding
quote. Hence, we cannot determine the trade sign by comparing the
traded price and the preceding midpoint price. This latter definition
of the trade sign was employed by Lee and
Ready~\cite{Lee1991}. Instead, we here define the trade signs
similarly to the tick rule of Holthausen, Leftwich and
Mayers~\cite{Holthausen1987}. They define the trade as
buyer--initiated (seller--initiated) if the trade is carried out at a
price above (below) the prior price. Zero tick trades are not
classified in general. The tick rule has an accuracy of $52.8\%$
\cite{Holthausen1987}. For our study, we
further develop this method: as our data has a one--second accuracy in
time, we consider the consecutive time intervals of length one second.
Let $t$ label such an interval and let $N(t)$ be the number of trades
in that interval. The individual trades carried out in this interval
are numbered $n=1,\ldots,N(t)$ and the corresponding prices are
$S(t;n)$.  We define the sign of the price change between consecutive
trades as
\begin{eqnarray}       
\varepsilon(t;n)=\left\{                  
\begin{array}{lll}    
\mathrm{sgn}\bigl(S(t;n)&-&S(t;n-1)\bigr)   \ ,  \\
                                      &&~~\mbox{if}~~S(t;n)\neq S(t;n-1), \\    
            \varepsilon(t;n-1) &&,~\mbox{otherwise}.
\end{array}           
\right.              
\end{eqnarray}
If two consecutive trades of the same trading direction together did
not exhaust all the available volume at the best price, the price of
both trades would be the same.  Thus, we set the trade sign equal to
the previous trade sign in this case. If there is more than one trade
in the interval denoted $t$, we average the corresponding trade
signs,
\begin{eqnarray}       
\varepsilon(t)=\left\{                  
\begin{array}{lll}    
\mathrm{sgn}
\left(\sum\limits_{n=1}^{N(t)}\varepsilon(t;n)\right) & \ , \quad & \mbox{if} \quad N(t)>0 \ , \\    
                                                0 & \ , \quad & \mbox{if} \quad N(t)= 0 \ ,
\end{array}           
\right.              
\end{eqnarray}     
which formally also includes the case $N(t)=1$. Consequently
$\varepsilon(t)=+1$ implies that the majority of trades in second $t$
was triggered by a market order to buy and, a
value $\varepsilon(t)=-1$ indicates the majority of sell market
orders. We have $\varepsilon(t)=0$, whenever trading did not take place in the time
interval $t$ or if there was a balance of buy and sell market
orders. In order to avoid overnight effects and any artifacts at the
opening and closing of the market, we consider only trades of the same
day from 9:40 to 15:50 New York local time.

\section{Response for pairs of stocks} 
\label{section3}

To study the mutual dependences between stocks, we consider sets of
two different stocks, to which we refer as pairs. We introduce a
response function as well as a trade sign correlator for such stock
pairs in Sects.~\ref{sec31} and Sect.~\ref{sec32}, respectively. After
an empirical analysis of these quantities, we discuss a certain noise
in Sect.~\ref{sec33}. In the sequel, all quantities referring to a
particular stock carry its indexed $i$,
quantities referring to a pair carry two such indices. We
consider eight pairs of the stocks listed in Table~\ref{table1}, four
within the same economic sector, four across different economic
sectors.

\subsection{Response functions} 
\label{sec31}

\begin{figure*}[htbp]
  \begin{center}
    \includegraphics[width=0.8\textwidth]{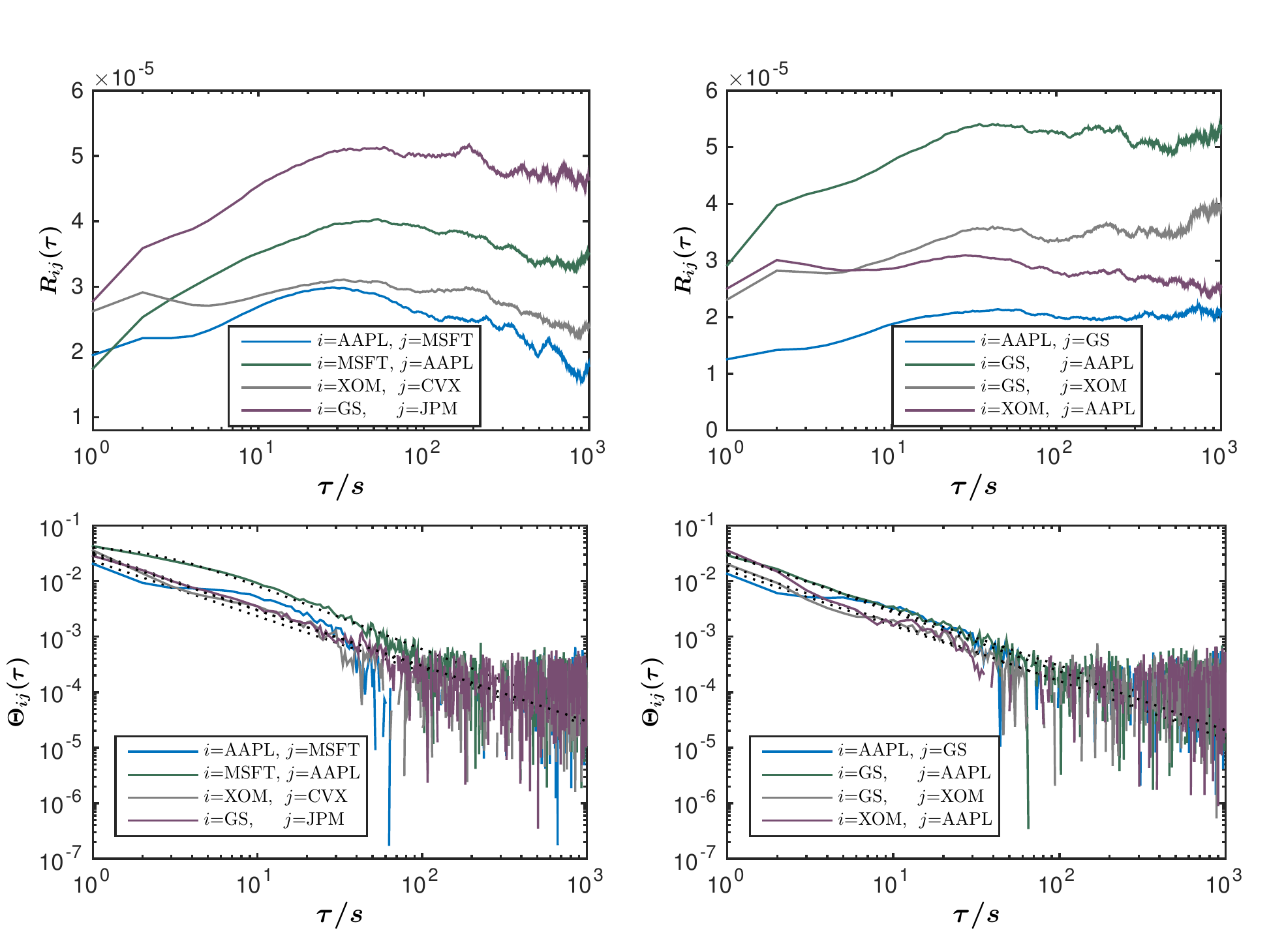}
  \end{center}
  \caption{ Response functions $R_{ij}(\tau)$ in 2008 versus time
        lag $\tau$ on a logarithmic scale (top left and right). Corresponding trade sign correlator $\Theta_{ij}(\tau)$ for different stock
    pairs on a doubly logarithmic scale, fits as black dotted lines (bottom, left and right). Stock pairs from the same economic sector (left), pairs of stocks from different sectors (right).}  
 \label{Fig.1}
\end{figure*}

To measure how a buy or sell order of stock with indexed $j$ at
time $t$ influences the prices of the stock $i$ at a later time
$t+\tau$, we introduce a new response function. We employ the
logarithmic price differences or log--returns for stock $i$ and time
lag $\tau$, defined via the midpoint prices $m_i(t)$,
\begin{equation}
r_i(t,\tau) \ = \ \log m_i(t+\tau) -\log m_i(t) \ = \ \log\frac{m_i(t+\tau)}{m_i(t)}
\label{eq0}
\end{equation}
at a given time $t$, keeping in mind the one--second accuracy.  To
acquire statistical significance, the response function is the time
average
\begin{equation}
R_{ij}(\tau) \ = \ \Bigl\langle r_i(t,\tau)\varepsilon_j(t)\Bigr\rangle _t \ 
\label{eq1}
\end{equation}
of the product of time--lagged returns and trade signs for stocks
$i$ and $j$, respectively. Our empirical results of this response
function for different stock pairs $(i,j)$ are shown in Fig.~\ref{Fig.1} versus the time lag.  In all cases, an increase to a
maximum is followed
by a decrease, \textit{i.e.}  the trend in the response is eventually
always reversed.  This does not depend on whether or not the pairs are
in the same economic sector or extend over two sectors. 

The stocks face similar systematic risks, leading to stronger response in the same sector than
across different sectors. However, strong responses for the stock pairs
from different sectors also exist, \textit{e.g.} for (GS, AAPL).
Apart from reasons specific for the stock pair considered, this might also be related to how investors assemble their portfolios. To disperse
the investment risks, the portfolios often comprise stocks from
different sectors since they are exposed to different economic risks
and are less correlated than stocks within the same sector. When investors buy or sell the stocks in their portfolios gradually, it may produce  responses and sign correlations in different stocks. We measure the strength of the sign correlation in Sect.~\ref{sec32}.

Fig.~\ref{Fig.1} shows that the response increases again after
decreasing back at large time lag $\tau$, \textit{i.e.}, $\tau$ close to 1000 s. We attribute this to the response noise that we introduce in Sect.~\ref{sec33}.

\subsection{Trade sign correlator}
\label{sec32}

The existence of sign correlations is the main reason that causes the
response in a single stock~\cite{Bouchaud2004}. For pairs of stocks, we
introduce the trade sign correlator
\begin{equation}
\Theta_{ij}(\tau) \ = \ \Bigl\langle \varepsilon_i(t+\tau)\varepsilon_j(t) \Bigr\rangle _t
\label{eq2}
\end{equation}
as a function of the time lag $\tau$.  As we demonstrate in Fig.~\ref{Fig.1}, there are a non--zero correlations across stocks.
It turns out that the empirical results can be fitted well by the
power law
\begin{equation}
  \Theta_{ij}(\tau)=\frac{\vartheta_{ij}}{\left(1+(\tau/\tau_{ij}^{(0)})^2\right)^{\gamma_{ij}/2}} \ .
\label{eq3}
\end{equation}
To estimate the error, we use the normalized $\chi_{ij}^2$ (see
App.~\ref{appB}). The parameters for the best fit as well as the $\chi_{ij}^2$ values for the analyzed
eight stock pairs are listed in Table~\ref{table2}.

\begin{table}[b]
  \caption{Fit parameters and normalized $\chi_{ij}^2$ for the trade sign correlators.} 
\begin{center}
\begin{tabular}{llcccc} 
\hline
\hline
stock $i$~~~~&stock $j$ ~& ~~~$\vartheta_{ij}$~~~ &~~~ $\tau_{ij}^{(0)}$~~~&~~~$\gamma_{ij}$ ~~~&$\chi_{ij}^2$ \\
				&					&							&		 [ s ] 					&							&~($\times10^{-6}$)\\
\hline
AAPL		&MSFT 			&0.46					&0.05						& 1.00					 &0.23 \\
MSFT		&AAPL			&0.04					& 2.34						&1.15 					 &0.10\\
XOM			& CVX			&0.61					&0.06						&1.04 					 &0.07\\
GS			&JPM			& 0.45					&0.07						&1.00					&0.04\\
AAPL		&GS				&0.46					&0.03						&1.00 					&0.11\\
GS			&AAPL			&0.49 					& 0.06						&1.00					&0.05 \\
GS			&XOM			&0.61					&0.04						&1.04					& 0.04\\
XOM			&AAPL			&1.18					& 0.03						&1.06					&0.13\\
\hline
\hline
\end{tabular}
\end{center}
\vspace*{-0.3cm} 
\label{table2}
\end{table}

In contrast to the sign correlation in one single stock, the stock
pair correlations exhibit short memory with exponents $\gamma_{ij}\geq
1$ rather than long memory, which usually is defined as corresponding
to exponents smaller than unity~\cite{Beran1994}. This indicates that
the price change of one stock responding to the trades of another
stock only persist for shorter times, and the response reverses at
relatively small time lags $\tau$. We notice the large fluctuations of
the trade sign correlator at larger lags $\tau$. They are partly due
to the decrease of the response signal, but also to the limited
statistics. The larger the time lag $\tau$, the larger is the overlap of the lag $\tau$ for different times $t$. When averaging the sign correlation over every second $t$ with large $\tau$, the result has poor statistics.

\subsection{Response noise} 
\label{sec33}

\begin{figure}[t]
  \begin{center}
  \includegraphics[width=0.49\textwidth]{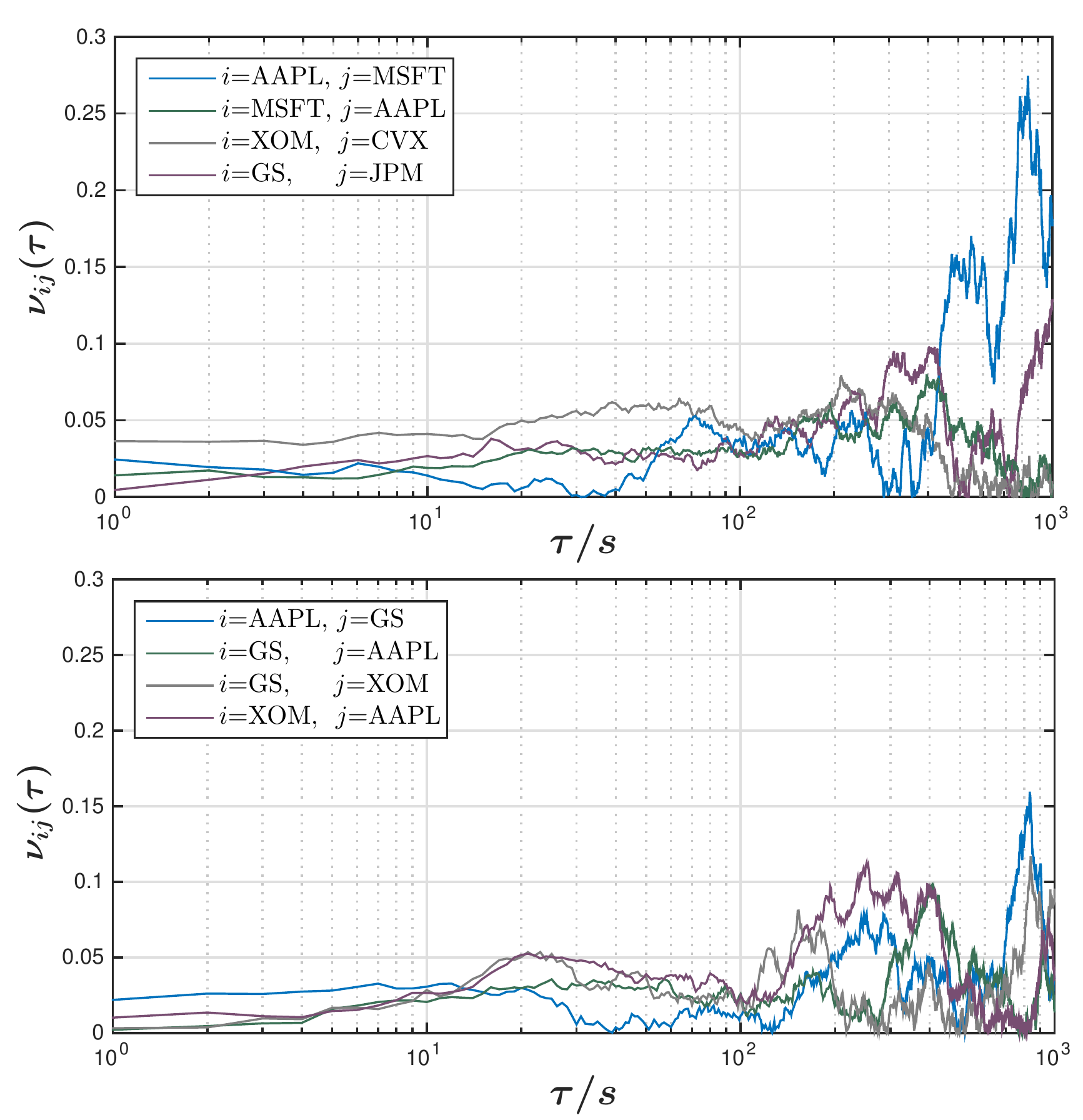}
   \end{center}
  \caption{Response noise $\nu_{ij}(\tau)$ for stock different pairs
    during the year 2008 versus the time lag $\tau$ measured on
    a logarithmic scale. Stock pairs from the same economic sector
    (top), pairs of stocks from different sectors (bottom).} 
 \label{Fig.2}
\end{figure}

As pointed out above, the response functions and the sign correlators
strongly fluctuate during the decay. Here, we address this point by
introducing the response noise $\nu_{ij}(\tau)$ as an estimator: We
determine the number $T_{ij}^{(c)}$ common to stocks $i$ and $j$ in which
trading took place.  We label these days with a running integer number
and separate our data into two sets, for days with even and odd
numbers, respectively. We work out the corresponding response
functions $R_{ij}^{(k)}(\tau)$ with $k=1,2$ for the averages over odd or
even days.  Each of these two functions should be very close to the
response function $R_{ij}(\tau)$ averaged over all days. Thus, we introduce
a response noise as some kind of normalized Euclidian distance
\begin{equation}
\nu_{ij}(\tau) \ = \ \frac{1}{|R_{ij}(\tau)|}
            \sqrt{\frac{1}{2}\sum^{2}_{k=1}\left(R_{ij}^{(k)}(\tau)-R_{ij}(\tau)\right)^2} 
\label{eq4a}
\end{equation}
for each value of the time lag $\tau$. In Fig.~\ref{Fig.2} we present
the empirical results for the response noise during the year 2008.
Obvious, most stock pairs do not suffer from large response noise for
time lags smaller than about 120 seconds. During this period, the
noise lies below a value of about 0.06. With increasing time lag, the
noise becomes much strong, indicating unstable response. The largest
noise reaches values of more than 0.25 for lags tending towards 1000
seconds. This is the reason why some stock pairs show upwards trends
after reversing back. As the sign correlator weakens in the regime of
large time lag, other factors dominate leading to the large response
fluctuations. Limited statistics blurs the picture, since there are
only 22200 seconds of effective trading time in each trading day. This
clearly demonstrates that, when looking at the response of a stock
pair, the lags considered must not be too large to obtain meaningful
results. In the sequel, we overcome the problems related to the
limited statistics, by further averaging the response functions over
different stock pairs.

\section{Market response} 
\label{section4}

The response functions and the trade sign correlators we considered up
to now give us a kind of microscopic information for stock pairs. It
is equally important to investigate how the trading of individual
stocks influences the market as a whole. In a first step, we tackle
this question by introducing the market response as the matrix
$\rho(\tau)$ whose entries are the normalized response functions at a
given time lag,
\begin{equation}
\rho_{ij}(\tau) \ = \ \frac{R_{ij}(\tau)}{\textrm{max\,}(|R_{ij}(\tau)|)} \ ,
\label{eq4b}
\end{equation}
where the denominator is the maximum over all stock pairs $(i, j)$ for fixed $\tau$.
This object is reminiscent of, but should not be mixed up with a
correlation matrix. Importantly, the matrix of the market response is
not symmetric, $R_{ij}(\tau)\neq R_{ji}(\tau)$, as to different
quantities, the returns and the trade signs, enter the
definition Eq.~\eqref{eq1}. Furthermore, the market response reveals information about the time evolution. 

\begin{figure}[b]
  \begin{center}
    \includegraphics[width=0.49\textwidth]{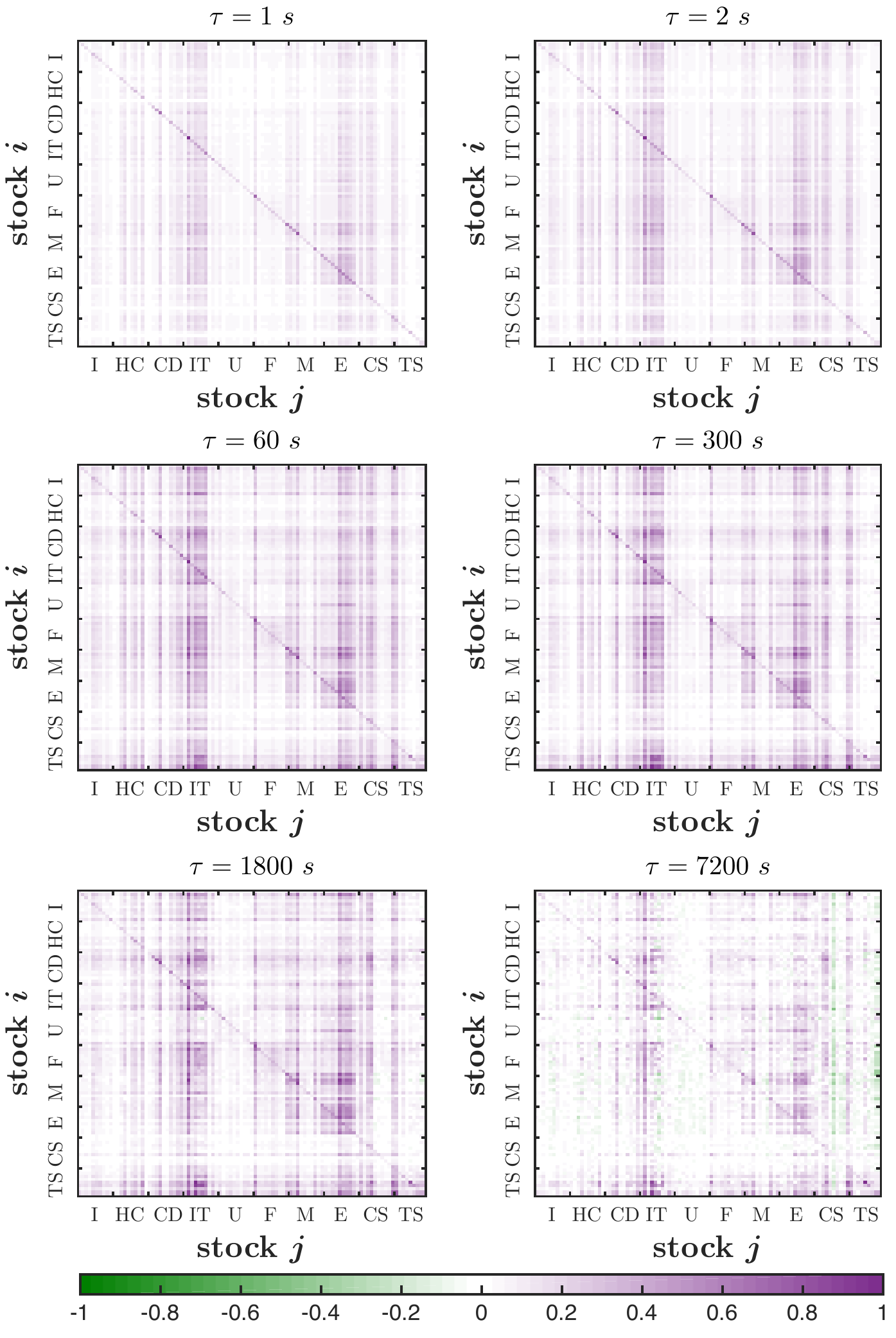}
  \end{center}
  \caption{Matrices of market response with entries $\rho_{ij}(\tau)$
    for $i,j=1,\ldots,99$ at different time lags $\tau=1, 2, 60, 300,
    1800, 7200$~s in the year 2008. The stocks pairs $(i,j)$ belong to
    the sectors industrials (I), health care (HC), consumer
    discretionary (CD), information technology (IT), utilities (U),
    financials (F), materials (M), energy (E), consumer staples (CS),
    and telecommunications services (TS).}
 \label{Fig.3}
\end{figure}

Our empirical analysis is depicted in Fig.~\ref{Fig.3} for a market with 99 stocks (see App.~\ref{appA}).
In Fig.~\ref{Fig.3} we show the $99\times99$ matrices of the market response for different
time lags $\tau=1, 2, 60, 300, 1800, 7200$~s in the year 2008.  The
diagonal strip is simply the response of the stock to itself.  In
general, the price change of one stock is always affected by the
trading of all others, and \textit{vice versa}.  The stocks are
ordered according to the economic sectors, and the six matrices of the
market response in Fig.\ref{Fig.3} feature striking patterns of strips
which can be associated with these sectors. For example, the
information technology (IT) sector produces a visibly strong strip
over almost all other sectors. This effect is quite stable over
time. It is worth mentioning that the price responses vary from sector
to sector. For example, the energy (E) also has strong response
but utilities (U) have weaker response.

\begin{figure}[t]
  \begin{center}
    \includegraphics[width=0.49\textwidth]{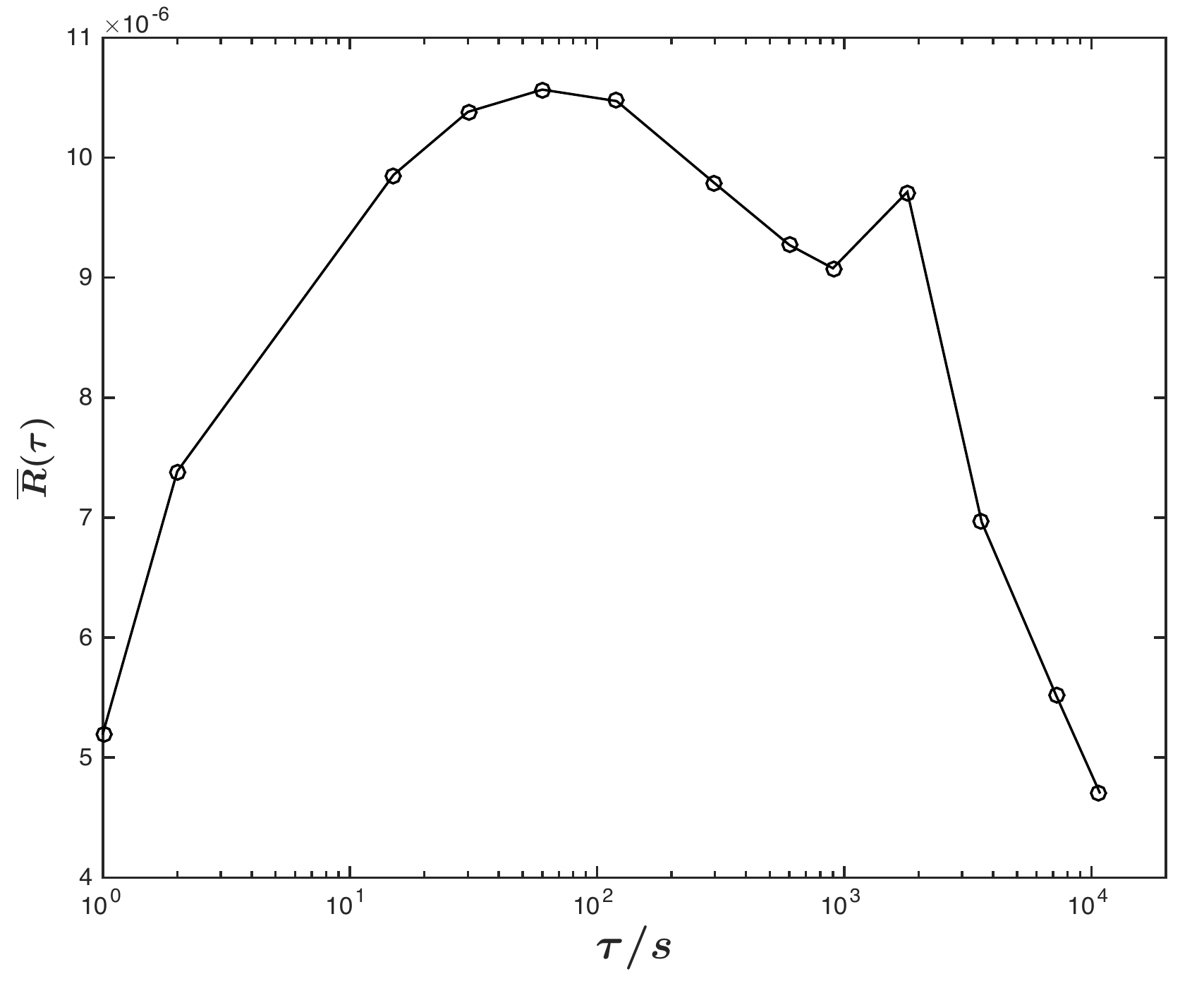}
  \end{center}
  \caption{Average response function $\overline{R}(\tau)$ for the whole market in 2008 versus time
        lag $\tau$ on a logarithmic scale. }
 \label{Fig.4}
\end{figure}

As seen in Fig.~\ref{Fig.3}, the market response is mainly positive up
to time lags of about $\tau=7200$~s, while negative responses show up
later. According to the Efficient Market Hypothesis
(EMH)~\cite{Fama1970}, the price encodes all available information,
implying that arbitrage opportunities do not exist. As the response
functions measure the price changes caused by trade signs, they should
according to the EMH be zero, for one single stock as well as across
different stocks. However, the empirical response functions for one
single stock already demonstrated that the existence of non--zero response value~\cite{Bouchaud2004}. Here, we go beyond this.  The
non--zero response functions across different stocks that we find show
the lack of efficiency for the market as a whole. Our results allow us
to extend the interpretation put forward in Ref.~\cite{Bouchaud2004}.
The impact of trades on the prices is transient with lag--dependent
characteristics. The trends due to potentially information--driven
trading by one or several of the market participants will be reversed
by others who act as arbitrageurs until a state that is compatible
with the EMH is reached again. This process involves the market as a
whole, not only the stock that is traded because of potential
information. The market needs more time to respond to all the
potential information before becoming efficient again. We showed in
Sect.~\ref{section3} that the responses fluctuate at large time lags
due and interpreted this as a noise effect. However, for the whole
market, these fluctuations are washed out by a self--averaging process
amounting to
\begin{equation}
\overline{R}(\tau) \ = \ \langle\langle R_{ij}(\tau)\rangle _{j}\rangle _{i} \ ,
\label{eq4c}
\end{equation}
where $i=j$ is excluded. The average response $\overline{R}(\tau)$ for the whole market is shown in Fig.~\ref{Fig.4} versus the time lag $\tau$. An increase of $\overline{R}(\tau)$ is followed by a decrease similar to the response for a stock pair. For the whole market, however, the decay takes longer and is observable for time scales up to three hours. We conclude that the market impact is transient. Efficiency is restored only after these decay processes.

\section{Average response functions}
\label{section5}

We define passive and active response functions in Sect.~\ref{sec51},
and then analyze them for the whole market and for economic sectors in
Sect.~\ref{sec52} and~\ref{sec53}.

\subsection{Passive and active response functions} 
\label{sec51}

As seen in Fig.~\ref{Fig.1}, the responses of one stock to other
stocks varies quite a bit. For example, the three stock pairs
$(\textrm{GS},j)$ with $j=\textrm{JPM},\textrm{AAPL},\textrm{XOM}$,
show different responses dependent on $\tau$. There are similar
differences for the three stock pairs $(i,\textrm{AAPL})$ with
$i=\textrm{MSFT},\textrm{GS}, \textrm{XOM}$. Moreover, the market
response displayed in Fig.~\ref{Fig.3}, quantifies the mutual impacts.
Typically, a given stock is related to several or many others by
trading.  As already mentioned, that is partly due to the grouping of
investments in portfolios, but there are may other reasons for the
mutual impact.  Suppose, for example, a trader who considers AAPL as
presently underpriced and likely to raise in the near future. To buy
many shares of AAPL he might use the profit from selling other stocks.
If many others act correspondingly, an impact results: buying
(selling) AAPL affects the stocks which are sold (bought). By
discussing this scenario, we want to motivate that averaging the
response functions over different stocks that are paired with the same
stock can yield interesting new observations. Furthermore, such
averages will also to some extent smoothen the drastic fluctuations of
the sign correlations at large time lags, \textit{cf.}, Fig.~\ref{Fig.1}, and reduce the response noise, \textit{cf.},
Fig.~\ref{Fig.2}. As the definition Eq.~\eqref{eq1} of the response
functions is not symmetric with respect to the indices, we can perform
two conceptually different averages,
\begin{equation}
R_i^{(p)}(\tau)=\left\langle R_{ij}(\tau)\right\rangle_{j}
\quad \textrm{and} \quad
R_j^{(a)}(\tau)=\left\langle R_{ij}(\tau)\right\rangle_{i},
\label{eq4d}
\end{equation}
to which we refer as passive and active response
functions. Importantly, the self--responses for $(i,i)$ or $(j,j)$ are
excluded in these averages.  The passive response function
$R_i^{(p)}(\tau)$ measures, how the price of stock $i$ changes due to
the trading of all other stocks, while the active response function
$R_j^{(a)}(\tau)$ quantifies which effect the trading of stock $j$ has
on the prices of all other stocks. Correspondingly, we also introduce
\begin{equation}
\Theta_i^{(p)}(\tau) = \left\langle \Theta_{ij}(\tau)\right\rangle_{j}
\quad \textrm{and} \quad
\Theta_j^{(a)}(\tau) = \left\langle \Theta_{ij}(\tau)\right\rangle_{i} ,
\label{eq4e}
\end{equation}
as passive and active trade sign correlators. We notice that they are
not symmetric either, because the time lag only enters the trade sign
with index $i$.

\subsection{Average responses of an individual stock to the market}
\label{sec52}

\begin{figure*}[htbp]
  \begin{center}
    \includegraphics[width=0.8\textwidth]{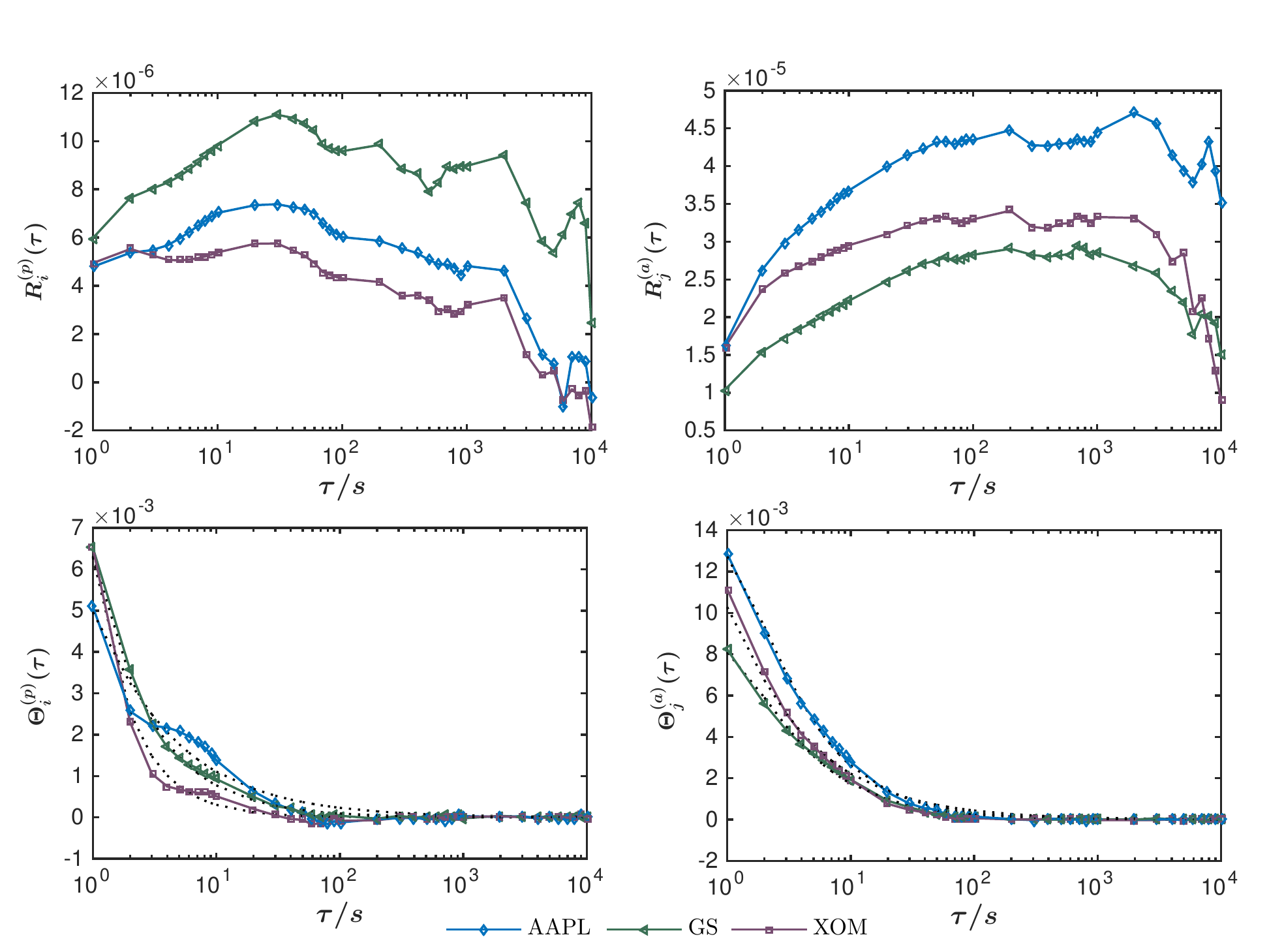}
  \end{center}
  \caption{Passive and active response functions $R_i^{(p)}(\tau)$ and
    $R_j^{(a)}(\tau)$ for $i,j=\textrm{AAPL},\textrm{GS},\textrm{XOM}$
    in the year 2008 versus time lag $\tau$ on a logarithmic scale (top
    left and right). Corresponding passive and active trade sign
    correlators $\Theta_i^{(p)}(\tau)$ and $\Theta_j^{(a)}(\tau)$,
    fits as black dotted lines (bottom, left and right).}
 \label{Fig.5}
\end{figure*}

We carry out the empirical analysis for the stocks AAPL, GS, XOM by
averaging their response functions over other 495 stocks in the
S$\&$P500 index. The results for passive and active response functions
as well as the corresponding passive and active trade sign correlators are presented 
in Fig.~\ref{Fig.5}. We checked that these empirical results are
similar to those from averaging over the other 98 stocks seen in
App.~\ref{appA}. To facilitate the computation, we
only calculate the average response values at several time lags, at the marked positions shown in Fig.~\ref{Fig.5}, rather than at every second as in Sect.~\ref{section3}.

The passive and active response functions clearly show different
behaviors. The passive response reverses faster than the active one. It only persists dozens of seconds and then reverses to drop
down quickly with sizeable volatility. In contrast, the active
response reverses at time lags of some hundreds of seconds and the
\begin{figure}[htbp]
  \begin{center}
    \includegraphics[width=0.5\textwidth]{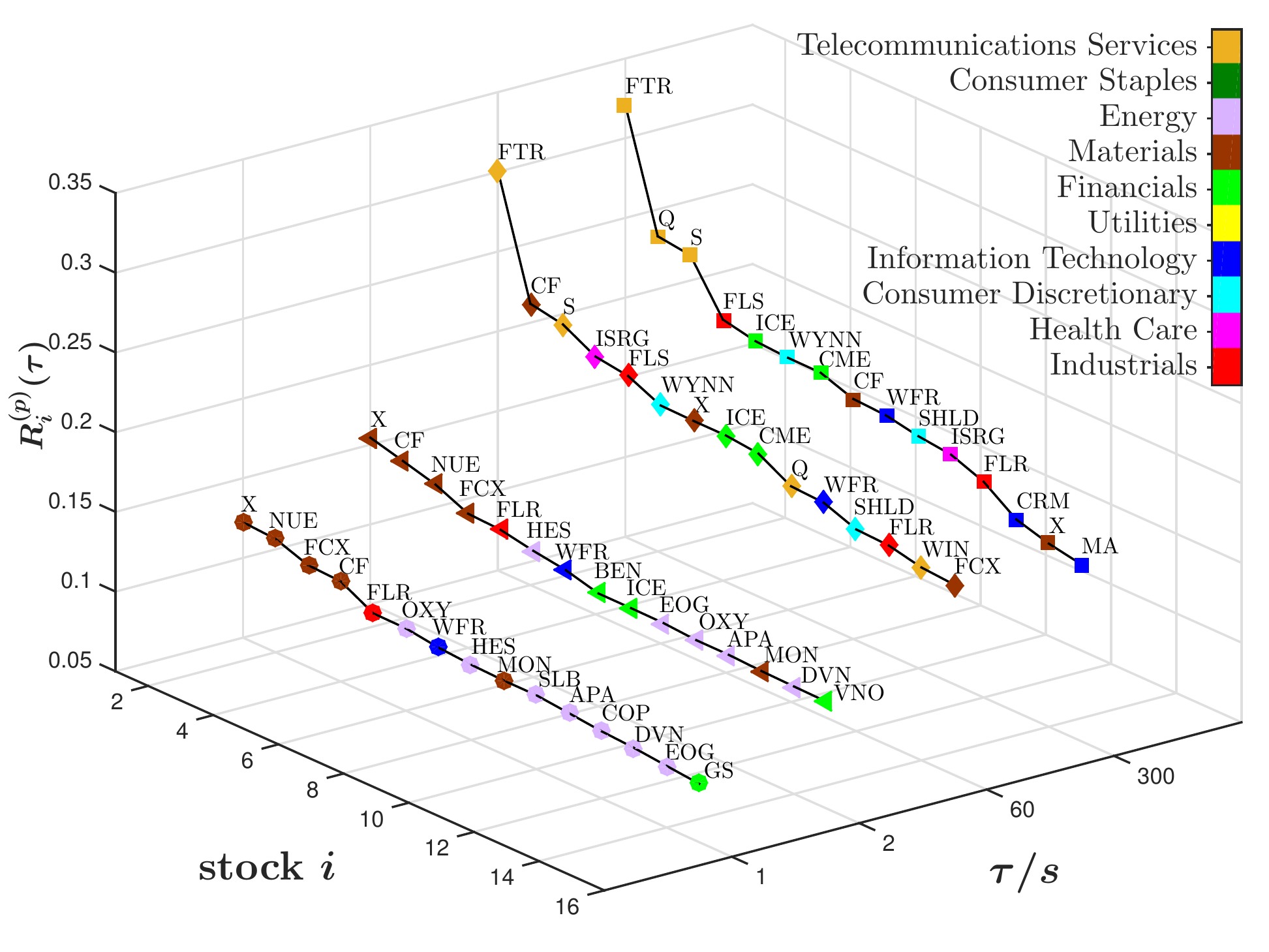}
    \includegraphics[width=0.5\textwidth]{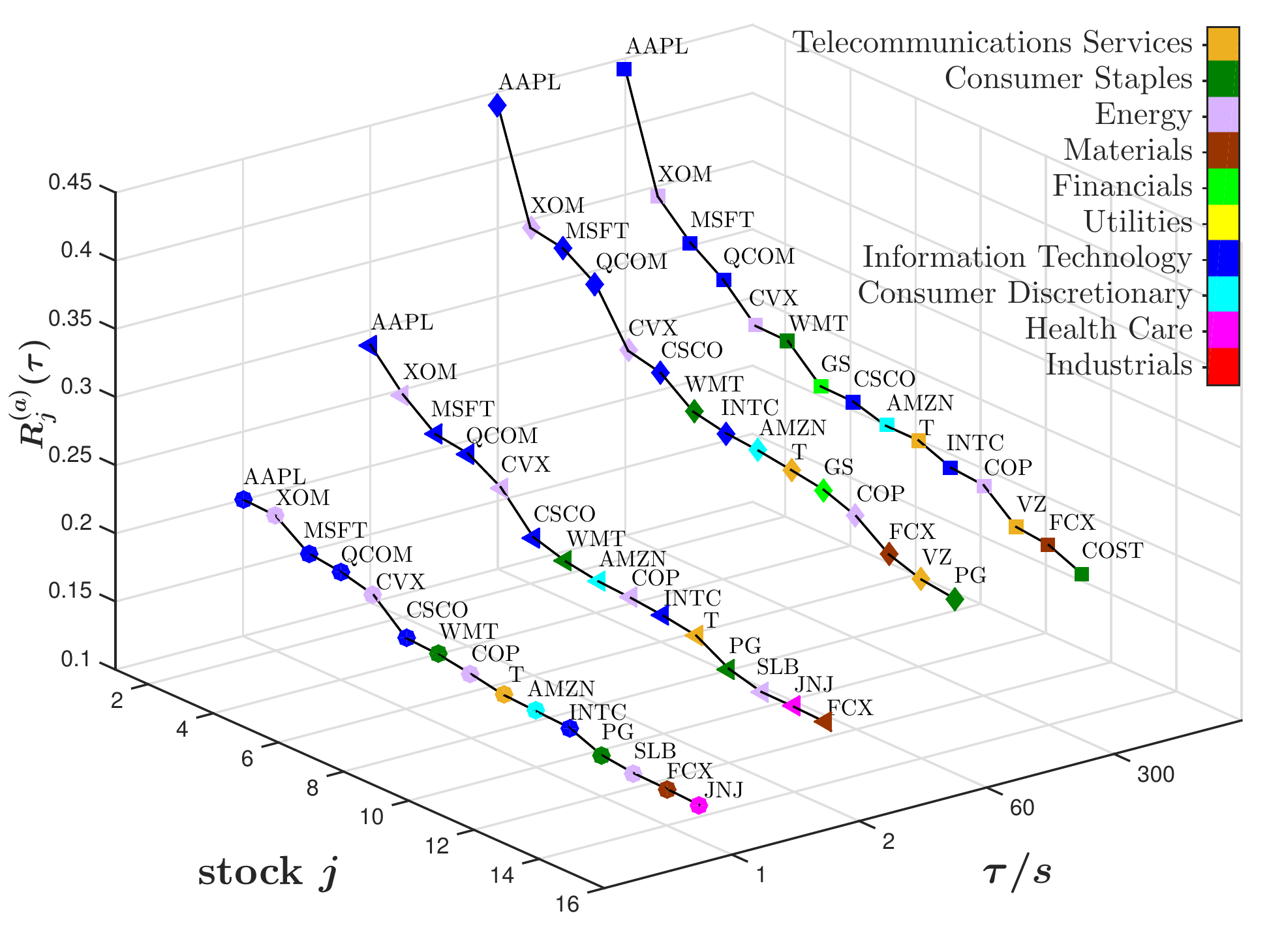}
  \end{center}
  \caption{The first fifteen stocks with strongest passive (top) and
    active (bottom) response functions $R_i^{(p)}(\tau)$ and
    $R_j^{(a)}(\tau)$ versus stock index $i$ or $j$ and time lags
    $\tau=1$~s ($\circ$), 2~s ($\triangleleft$), 60~s ($\diamond$),
    and 300~s ({$\Box$}). The ordinates of top and bottom graphs extend over different intervals.}
 \label{Fig.6}
\end{figure}
price changes slowly. An obvious reason for this difference is the
easier detectability of price changes in one stock than of those
dispersed over different stocks. Again, with our results, we can
extend the previous interpretations based on the study of single
stocks.  The passive response function reflects the price dynamics on
short time scales. When the price goes up, less market orders to buy
will be emitted and more limit orders to sell. Thus, the price
reverses~\cite{Hopman2007} without a need to evoke new information as
cause. Moreover, liquidity induced mean reversion attracts more
buyers, which motivates liquidity providers to raise the price again,
while the volatilities in this process of responding decline. Thus, we
conclude for the market as a whole that the mean reversion accentuates
the short--period price volatility, which is consistent with the
single--stock analysis~\cite{Handa1998,Bouchaud2010}. The active
response reflects the dispersion of the trade impacts over the prices
of different stocks. It is conceivable that this process takes longer
than compared with the time scales of the passive
response. Furthermore, this dispersion is accompanied by a spreading
out of the volatilities.

As visible in Fig.~\ref{Fig.5}, the active response is about five
times stronger than the passive one at the maximum values. 
The reason for different strength of passive and passive responses is the existence of strongly influential stocks. In Fig.~\ref{Fig.3}, we observe that the vertical stripes are much more pronounced than the horizontal ones. More specifically, there are groups of stocks which have a strong influence across most of the market, in particular, this is evident for the stocks in the IT sector. Consequently, the active response of these stocks averaged over the market shows a strong signal. To the passive response, however, fewer influenced stocks really contribute, which leads to reduced average. 

To identify the strongly influential and influenced stocks, we rank the 99 stocks in App.~\ref{appA} according to the numerical values of passive
and active response functions, normalized according to
Eq.~\eqref{eq4b}, at a given time lag $\tau$.  The first fifteen
stocks with strongest average response at $\tau=1, 2, 60, 300$~s
are shown in Fig.\ref{Fig.6}. As seen, FTR has stronger passive
responses for $\tau\geq 60$~s, implying that its price is more easily
impacted by the trades of other stocks. As a result, it has stronger passive response than active response, opposite to the cases of AAPL, GS and XOM in Fig.~\ref{Fig.5}. In contrast, AAPL, XOM, MSFT
and other stocks have stronger active responses, which means the
trades of these stocks are more likely to impact the prices of other
stocks. Interestingly, the ranking of individual stocks in the active responses 
looks similar at different time lags. This
matches the relatively stable response structure visible in
Fig.~\ref{Fig.3}.

To analyze the average trade sign correlators depicted in
Fig.~\ref{Fig.5}, we use the power law Eq.~\eqref{eq3} to fit the
empirical results. 
\begin{table}[b]
\caption{Fit parameters and $\chi^2$ of the average trade sign correlators.} 
\begin{center}
\begin{tabular}{clccc} 
\hline
\hline
Sign									&Parameters 								&\multicolumn{3}{c}{Stocks $i$, $j$}\\
\cline{3-5}
~correlators~					&~and errors~								& ~~AAPL~~ &~~~~GS~~~~ &~~XOM~~  \\
\hline
										&$\vartheta_{i}$							& 0.01 & 0.03 & 0.27 \\
$\Theta_i^{(p)}(\tau)$&$\tau_{i}^{(0)}$ [ s ] 				& 0.47 & 0.23 & 0.06 \\
										&$\gamma_{i}$ 							&  0.68 & 0.92 & 1.32  \\
										&$\chi^2_i$($\times10^{-7}$) &0.70 &  0.07 &   0.19 \\
										
										& $\vartheta_{j}$ 						& 0.02 & 0.01 & 0.02\\
$\Theta_j^{(a)}(\tau)$&$\tau_{j}^{(0)}$ [ s ] 				& 1.44 & 1.31 & 0.55\\
										&$\gamma_{j}$							& 0.90 & 0.85 & 0.71\\
										&$\chi^2_j$ ($\times10^{-7}$)& 0.28 & 0.18  & 1.06 \\										
\hline
\hline
\end{tabular}
\end{center}
\label{table3}
\vspace*{-0.3cm}
\end{table}
The fitting parameters and errors are shown in
Table~\ref{table3}. The remarkable result is that the volatile 
short memory of the individual correlators turns into a long memory
with exponents smaller than one after averaging. The only exception is
the passive trade sign correlator for the stock XOM. We thus infer
that the price changes caused by trade sign correlations in different
stocks can accumulate to persist over longer times.

\subsection{Average responses of an individual stock to economic sectors}
\label{sec53}

\begin{figure*}[htbp]
  \begin{center}
    \includegraphics[width=0.75\textwidth]{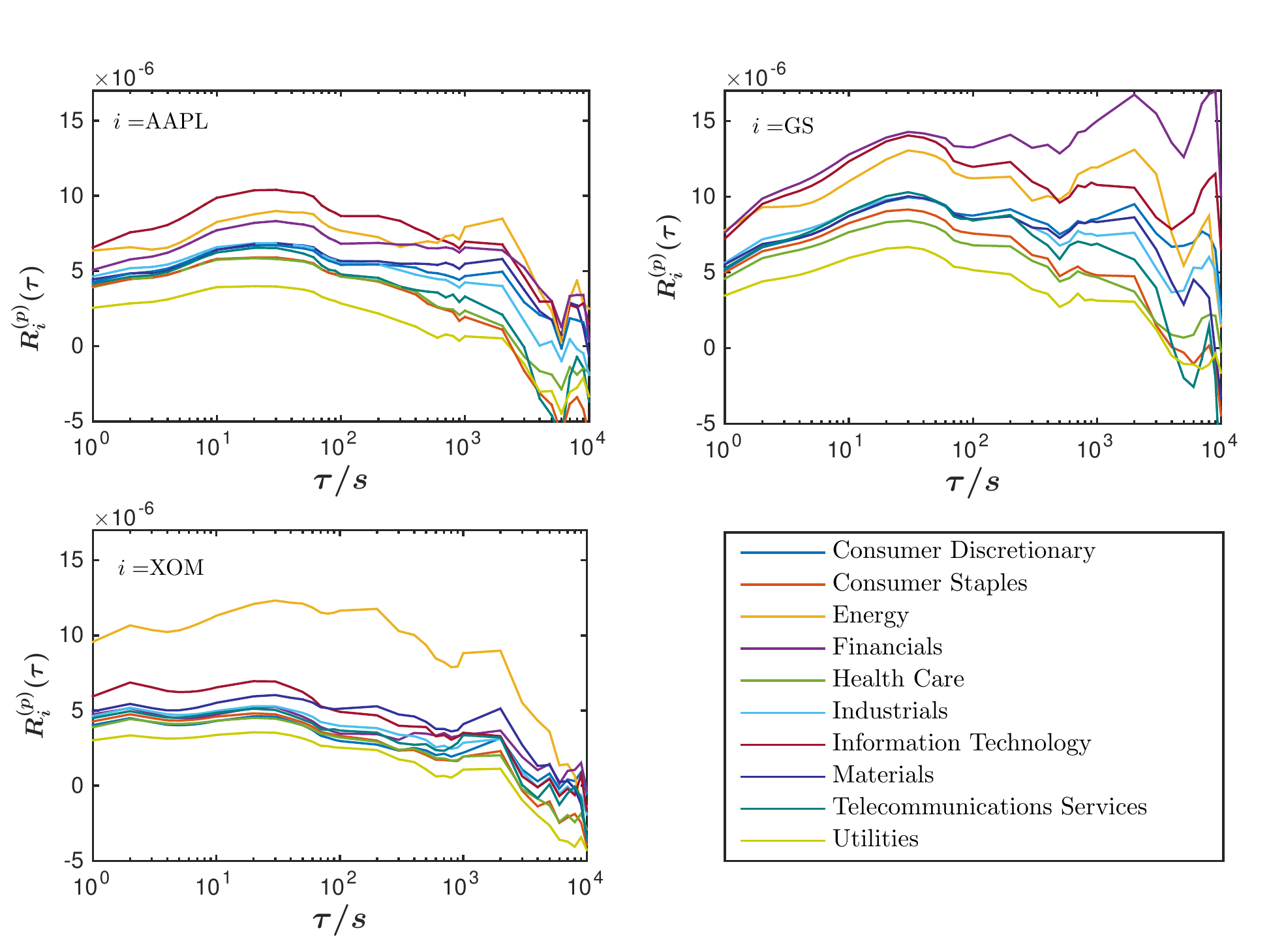}
  \end{center}
  \caption{Passive response functions $R_i^{(p)}(\tau)$ of the stocks
    $i=\textrm{AAPL},\textrm{GS},\textrm{XOM}$ to ten different
    economic sectors in the year 2008 versus time lag $\tau$ on a
    logarithmic scale.}
  \label{Fig.7}
\end{figure*}

\begin{figure*}[htbp]
  \begin{center}
    \includegraphics[width=0.75\textwidth]{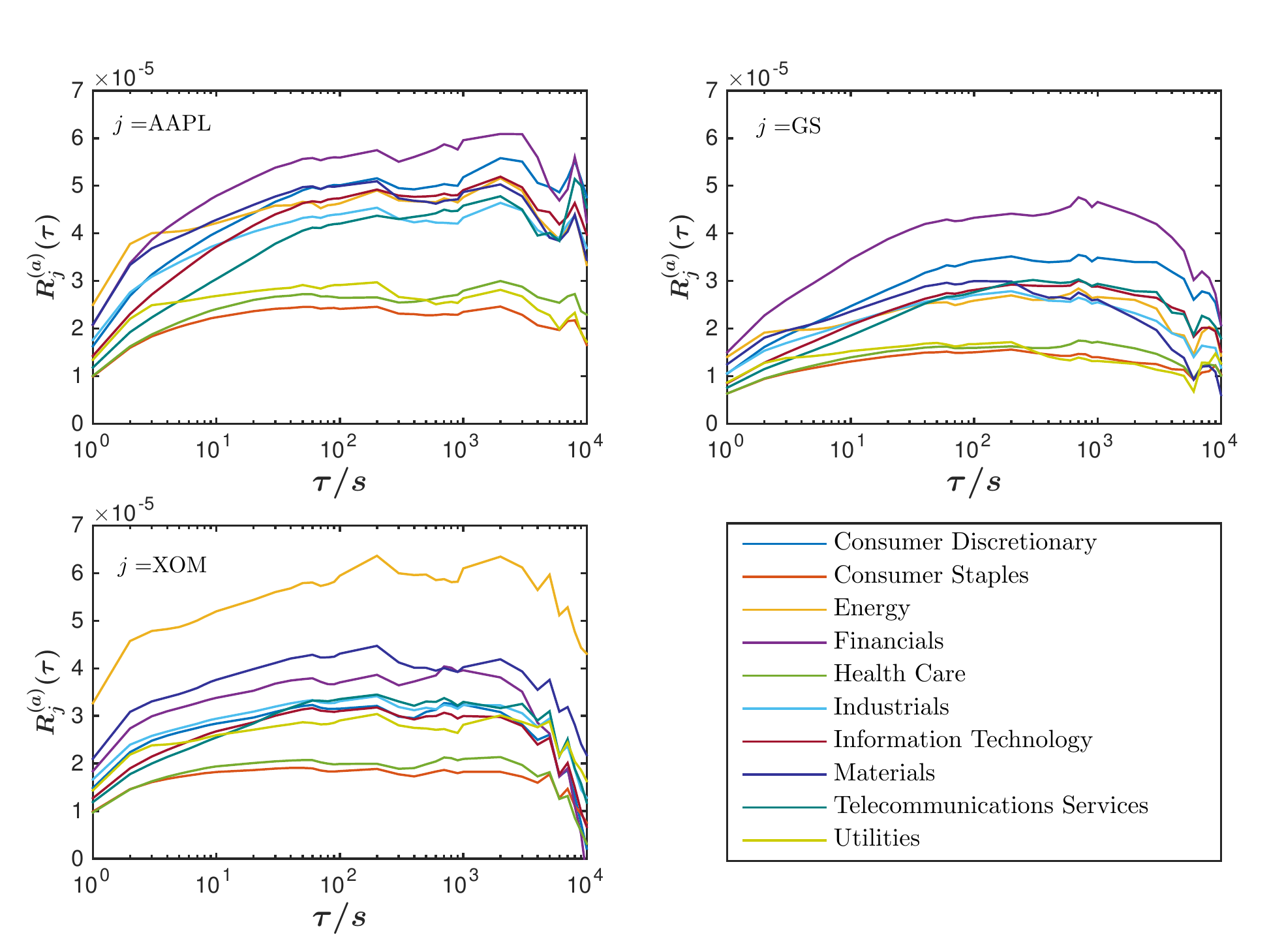}
  \end{center}
  \caption{Active response functions $R_j^{(a)}(\tau)$ of the stocks
    $j=\textrm{AAPL},\textrm{GS},\textrm{XOM}$ to ten different
    economic sectors in the year 2008 versus time lag $\tau$ on a
    logarithmic scale.}
 \label{Fig.8}
\end{figure*}

Another observation which can be made in Fig.~\ref{Fig.3} is that the
responses vary for different economic sectors. In other words, the
stocks from different sectors may produce different average responses
to a given stock. We calculate the average response of the stocks
AAPL, GS and XOM to ten economic sectors in the S$\&$P 500 index.  The
passive and active responses are displayed in Figs.~\ref{Fig.7}
and~\ref{Fig.8}, respectively. Clear differences are seen.

Regarding the passive responses, the prices of the three stocks
considered are all affected by the trades within their own sectors,
especially XOM, which is not surprising due to common economic
effects. Moreover, the price of the stock AAPL is also easily
influenced by both energy (E) and financials (F), similar observation
hold for the price of GS in relation to information technology (IT)
and energy (E).  Regarding the active response, the trades of AAPL and
GS have a significant impact on the prices of the stocks from
financials (F) significantly, but a lesser one utilities (U), health
care (HC), and consumer staples (CS). This might be due to the
stability of these sectors which serve the needs of daily life. The
trades of XOM are more likely to influence energy (E), but have a
lesser effect on health care (HC) and consumer staples (CS). This is
so because utilities (U) are economically more strongly coupled to
energy (E) than to health care (HC) and consumer staples (CS).

 \section{Comparisons between self-and cross-responses}
 \label{section6}

\begin{figure*}[htbp]
  \begin{center}
    \includegraphics[width=0.9\textwidth]{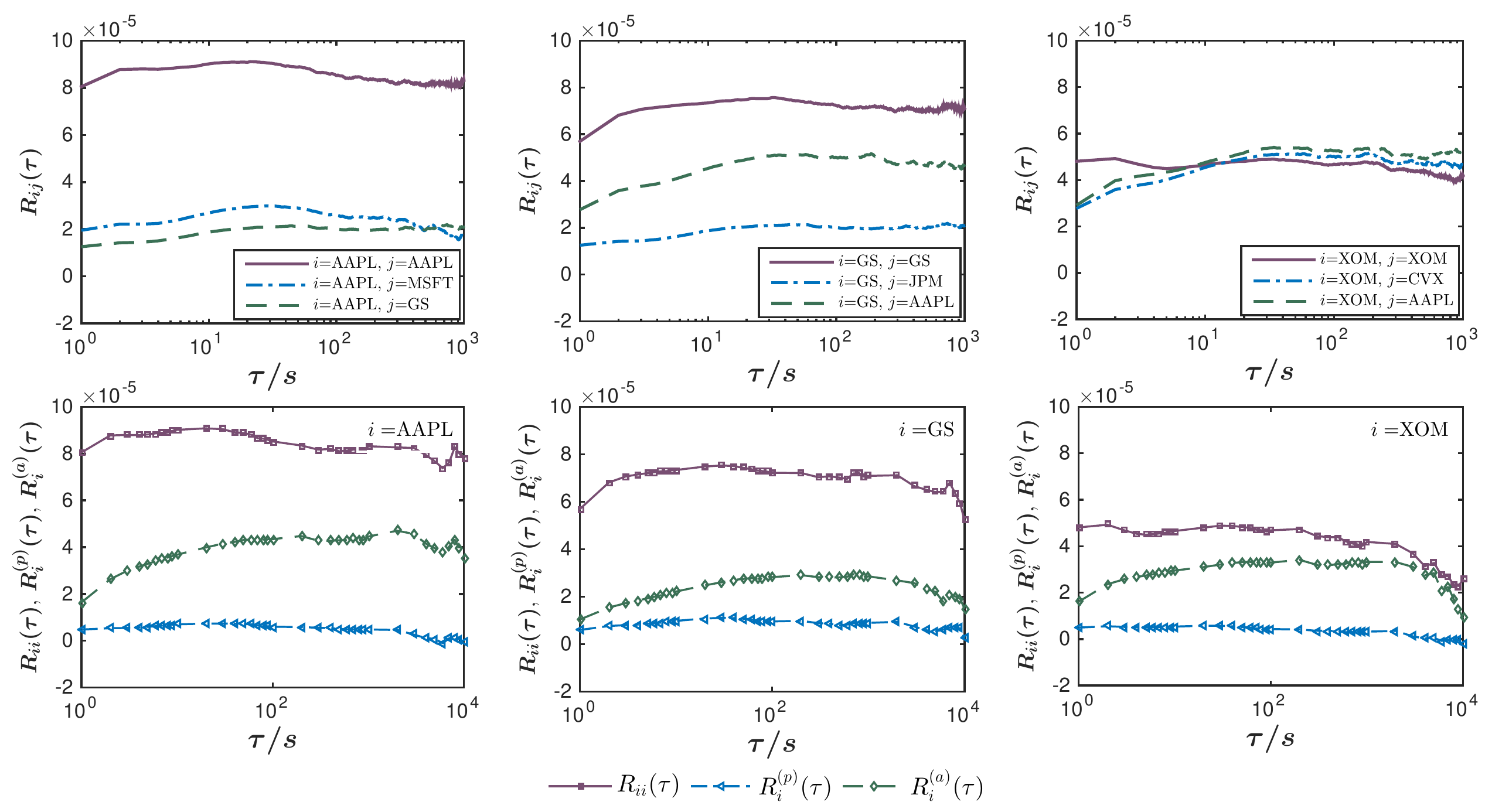}
  \end{center}
  \caption{Comparisons of the self-responses for AAPL, GS and XOM with cross-responses to different stocks (top) on a logarithmic scale. Comparisons of self-responses, passive responses, and active responses for the same stocks on a logarithmic scale (bottom).}
 \label{Fig.9}
\vspace*{2cm}
\end{figure*}

It is important to compare the self--response to the various cross-responses. In Fig.~\ref{Fig.9}, we show the self--responses for AAPL, GS and XOM together with cross-responses as well as together with the active and passive, \textit{i.e.} with averaged responses.

\begin{figure*}[htbp]
  \begin{center}
    \includegraphics[width=0.9\textwidth]{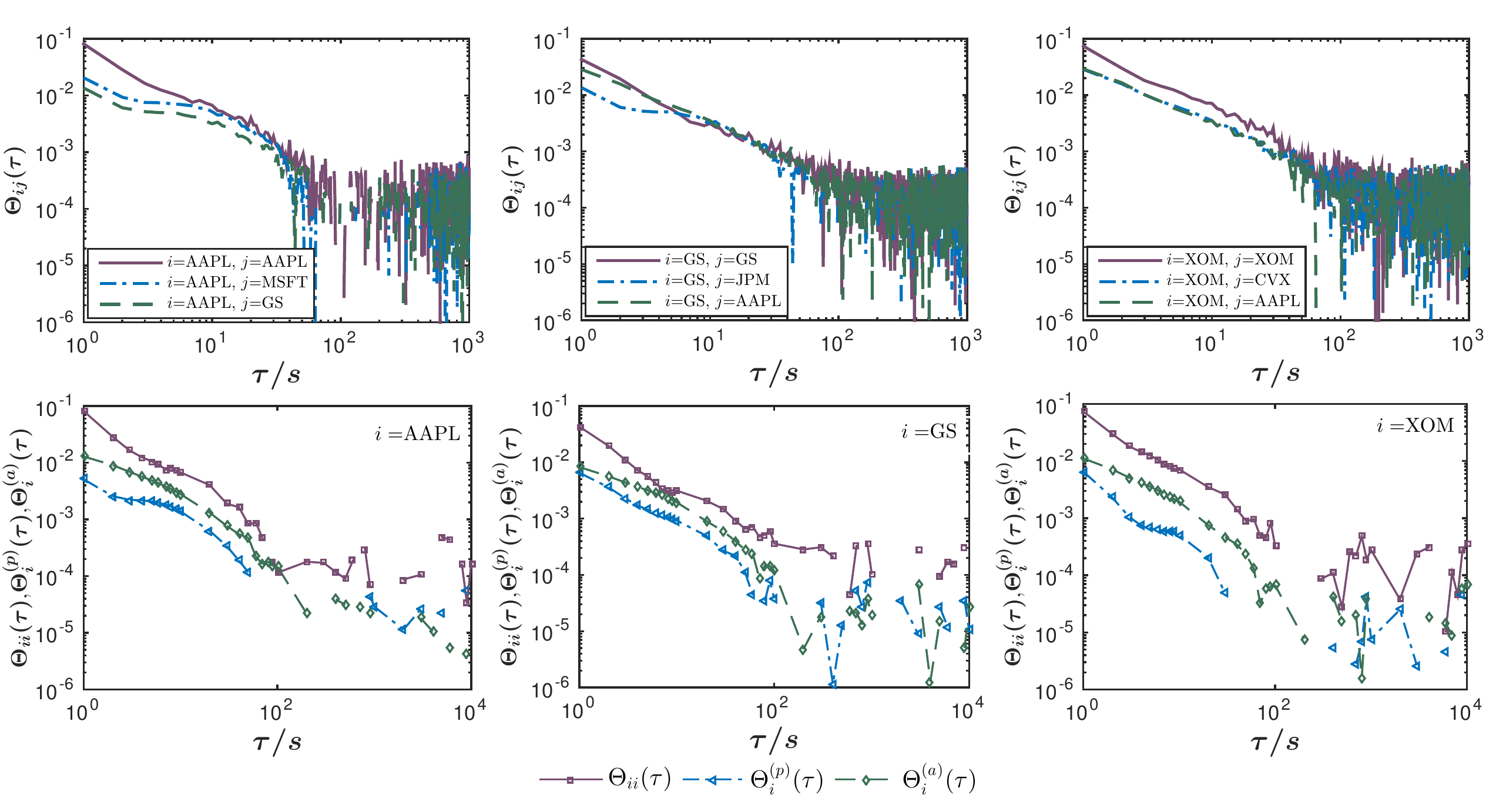}
  \end{center}
  \caption{Comparisons of the trade sign self-correlators for AAPL, GS and XOM with cross-correlators with different stocks on a doubly logarithmic scale (top). Comparisons of the self-correlators, passive correlators, and active correlators for the same stocks on a doubly logarithmic scale (bottom). The negative correlators are not shown in the graphs.}
 \label{Fig.10}
\end{figure*}

Typically, the self--response is stronger than the cross--response for two stocks due to the strong auto-correlation of trade signs. The example of XOM, however, shows that the cross--response can be stronger than the self--response. From Fig.~\ref{Fig.6}, we know the AAPL and CVX are influential stocks. Their trades are more likely to impact the price change of other stocks, such that the influenced stock, {\textit i.e.} XOM, responds to them more strongly than to itself.

The average responses are always weaker than the self--response, implying that cases such as XOM are rare. Due to the noise reduction we can follow the average responses over a longer time interval. On shorter scales of the time lag $\tau$, both, self-- and cross--responses should be considered when looking at individual stocks, but for investigating response stability, persistence or efficiency of the market as a whole, the average quantities give  useful information on longer scales.
 
Fig.~\ref{Fig.10} shows the trade sign self-- and cross--correlators for AAPL, GS and XOM corresponding to Fig.~\ref{Fig.9}. For the correlators between two stocks, the correlation differences are observable for time lags of less than 10 seconds. With the lag increasing, the self--correlators are close to the cross--correlators for AAPL and GS. The exception of XOM, however, is visible again in the sign correlators. In the research of the self--response~\cite{Bouchaud2004}, a 'bare' impact function with a power--law decay for a single trade is proposed to offset the amplification effect of sign self--correlations, which is derived from the correlation accumulation with the time lag increasing. Thus, both the impact function and the sign self--correlation mutually describe the price self--response. For XOM, the sign self--correlator is larger than the cross--correlators before the appearance of correlation fluctuations, but its cross--response is stronger than the self--response. It implies that the impact functions of the self-- and cross--responses are different. In other words, there are different impact mechanisms between the self--and cross--responses. The influential stocks, \textit{i.e.} AAPL and CVX, amplify the impact difference in the case of XOM. 

The average synthesizes the impacts of different stocks on individual stocks. The impacts cannot be observed directly. Instead, we can observe the decay of the average sign correlators. The passive and active correlators are always smaller than the self--correlators for AAPL, GS and XOM before the appearance of correlation fluctuations, which is consistent with the case of the responses. It implies that the synthesized impact functions of the individual stocks are more stable than the ones in each stock pair.

\section{Conclusions}  
\label{section7}

We extended the study of the responses of stock prices to trading from
individual stocks to a whole correlated market. We empirically
investigated the price responses to the trading of different stocks as
functions of the time lag. The response functions increase and then
reverse back. Thus, the impact of the trades on the prices 
appears to be transient. Pictorially speaking, the market needs time to react to the
distortion of efficiency caused by the potentially informed
traders. In this period of distortion, arbitrageurs drive the price to
a reversion and thereby help to restore market efficiency. The price
response is clearly related to the trade sign correlations for
different stocks. These correlators decays in a power--law fashion,
revealing a short--memory process with exponents larger than one for a
stock pair.

We also analyzed the market as a whole by setting up a matrix, the
market response, that collects the normalized information of all
response functions. Several characteristic features show up which are
visible in patterns having a remarkable stability in time. The market
response provides quantitative information about how the trading of one
stock affects the prices of other stocks, and how its own
price is influenced by the trades of other stocks stocks.

After this somehow microscopic view, we introduced average response
functions, a passive and an active one, measuring the average price
change of a given stock due to the trades of all others and the impact
of trading a specific stock on the average price change of the other
ones, respectively. Interestingly, the passive response reverses at a
relatively short time lag of dozens of seconds or so and then declines
rather quickly in a volatile way, while the active response reverses
at a longer time lag of some hundreds of seconds with less
volatility. This is so, because the price change in one stock easily
alerts the market participants. The dispersion over different stocks
makes it more difficult to detect an effect. We identified the
response noise as a criterion for the stability. The averaged
responses considerably reduce this noise, and make generic effects
visible.  We also introduced the corresponding active and passive
trade sign correlators. It is quite remarkable that the above
mentioned short memory turns into a long memory when averaged over
different stock pairs.

Some stocks dominate the price responses of the market in a very
stable fashion during the first 7200~s. By ranking the stocks, we
found that some influential stocks exhibit strong active response, but we also
identified an example for a strong passive response. Here, it is also
important that the responses in the market vary from sector to
sector. 

Last, the self--response is compared with the various cross--responses. On shorter scales of the time lag, both the self- and cross--responses should be considered for individual stocks. But on longer scales, the average responses of individual stocks give useful information for investigating the response stability, persistence or the efficiency of the market as a whole. On the other hand, the comparison of sign self-- and cross--correlators implies the existence of different impact mechanisms between self-- and cross--responses. By averaging the responses of individual stocks across the whole market, the impacts of individual stocks become stable.

\section*{Acknowledgements}

We thank T.A.~Schmitt, Y.~Stepanov, D. Chetalova, and D. Wagner for fruitful discussions. One of us (SW)
acknowledges financial support from the China Scholarship Council
(grant no. 201306890014).

\appendix

\section{Stocks used for analyzing the market response}
\label{appA}

We evaluated the market response for the 99 stocks from ten economic sectors: industrials (I), health care (HC), consumer	discretionary (CD), information technology (IT), utilities (U), financials (F), materials (M), energy (E), consumer staples (CS), and  telecommunications services (TS) as listed in Table~\ref{table4}. The acronym AMC in Table~\ref{table4} stands for averaged market capitalization.
\begin{table*}
\linespread{0.5} 
\caption{Information of 99 stocks from ten economic sectors} 
\begin{center}
\begin{footnotesize}
\begin{tabular}{llrcllr} 
\hline
\hline
\multicolumn{3}{l}{Industrials (I)} &~~~~~~~~& \multicolumn{3}{l}{Financials (F)}   \\
\cline{1-3}\cline{5-7}
Symbol	&Company							&AMC~ 		&		&Symbol	&Company								&AMC~		\\
\cline{1-3}\cline{5-7}
FLR		&Fluor Corp. (New)				&14414.4	&					&CME		&CME Group Inc.						&49222.9	\\
LMT		&Lockheed Martin Corp.		&12857.8	&					&GS			&Goldman Sachs Group			&21524.3	\\
FLS		&Flowserve Corporation		&12670.2	&					&ICE			&Intercontinental Exchange Inc.&14615.3	\\
PCP		&Precision Castparts			&12447.0	&					&AVB		&AvalonBay Communities			&11081.6	\\
LLL		&L-3 Communications 	Holdings&12170.8	&				 	&BEN		&Franklin Resources					&10966.2	\\
UNP		&Union Pacific						&11920.9	&					&BXP		&Boston Properties					&10893.0 	\\
BNI		&Burlington Northern Santa Fe C		&11837.5	&					&SPG		&Simon Property 	Group  Inc	 &10862.4	\\
FDX		&FedEx Corporation				&10574.7	&					&VNO		&Vornado Realty Trust				&10802.3	\\
GWW	&Grainger (W.W.) Inc.			&10416.8	&					&PSA		&Public Storage						&10147.9	\\
GD		&General Dynamics				&10035.6	&					&MTB		&M$\&$T Bank Corp.				&9920.2		\\
\cline{1-3}\cline{5-7}
\\
 \multicolumn{3}{l}{Health Care (HC)} &~& \multicolumn{3}{l}{Materials (M)}   \\
\cline{1-3}\cline{5-7}
Symbol	&Company							&AMC~ 		&					&Symbol	&Company								&AMC~		\\
\cline{1-3}\cline{5-7}
ISRG		&Intuitive Surgical Inc.			&31355.9	&	     			&X			&United States Steel Corp.		&15937.7	\\
BCR		&Bard (C.R.) Inc.					&11362.7	&	  				&MON		&Monsanto Co.							&14662.6	\\
BDX		&Becton  Dickinson				&10298.4	&					&CF			&CF Industries Holdings Inc		&14075.5	\\
GENZ	&Genzyme Corp.					&9728.8		&					&FCX		&Freeport-McMoran Cp $\&$ Gld &11735.7\\
JNJ		&Johnson $\&$ Johnson		&9682.6		&					&APD		&Air Products $\&$ Chemicals		&10246.4\\
LH		&Laboratory Corp. of America Holding		&9035.7&	 &PX			&Praxair  Inc.			&10234.5	\\
ESRX		&Express Scripts					&8864.6		&					&VMC		&Vulcan Materials						&8700.4		\\
CELG	&Celgene Corp.					&8783.1		&					&ROH		&Rohm $\&$ Haas					&8527.1		\\
ZMH		&Zimmer Holdings				&8681.7		&					&NUE		&Nucor Corp.							&7997.4		\\
AMGN	&Amgen								&	8543.0		&					&PPG		&PPG Industries						&7336.7		\\
\cline{1-3}\cline{5-7}
\\
\multicolumn{3}{l}{Consumer	Discretionary (CD)} &~& \multicolumn{3}{l}{Energy (E)}   \\
\cline{1-3}\cline{5-7}
Symbol	&Company							&AMC~ 		&					&Symbol	&Company								&AMC~		\\
\cline{1-3}\cline{5-7}
WPO		&Washington Post				&61856.1	&	 				&RIG			&Transocean Inc. (New)			&16409.5	\\
AZO		&AutoZone Inc.					&14463.7	&	    			&APA		&Apache Corp.							&13981.9	\\
SHLD	&Sears Holdings Corporation&11759.2	&					&EOG		&EOG Resources						&13095.0	\\
WYNN	&Wynn Resorts Ltd.				&11507.9	&					&DVN		&Devon Energy Corp.				&12499.7	\\
AMZN	&Amazon Corp.					&10939.2	&					&HES		&Hess Corporation					&11990.4	\\
WHR		&Whirlpool Corp.					&9501.9		&					&XOM		&Exxon Mobil Corp.					&11460.3	\\
VFC		&V.F. Corp.							&9051.2		&					&SLB		&Schlumberger Ltd.					&11241.1	\\
APOL	&Apollo Group						&8495.8		&					&CVX		&Chevron Corp.						&11100.0	\\
NKE		&NIKE Inc.							&8149.5		&					&COP		&ConocoPhillips						&10215.3	\\
MCD		&McDonald's Corp.				&8025.6		&					&OXY		&Occidental Petroleum				&9758.4		\\
\cline{1-3}\cline{5-7}
\\
\multicolumn{3}{l}{Information Technology (IT)} &~& \multicolumn{3}{l}{Consumer Staples (CS)}   \\
\cline{1-3}\cline{5-7}
Symbol	&Company							&AMC~ 		&		&Symbol	&Company								&AMC~		\\
\cline{1-3}\cline{5-7}
GOOG		&Google Inc.							&62971.6	&	  				&BUD		&Anheuser-Busch						&9780.6		\\
MA			&Mastercard Inc.					&28287.8	&					&PG			&Procter $\&$ Gamble				&9711.5		\\
AAPL		&Apple Inc.							&22104.1	&					&CL			&Colgate-Palmolive					&9549.2		\\
IBM			&International Bus. Machines&15424.9	&					&COST		&Costco Co.								&9545.9		\\
MSFT		&Microsoft Corp.					&10845.1	& 					&WMT		&Wal-Mart Stores						&9325.7		\\
CSCO		&Cisco Systems					&8731.4		&					&PEP		&PepsiCo Inc.							&9180.7		\\
INTC			&Intel Corp.							&8385.8		&					&LO			&Lorillard Inc.							&8919.0		\\
QCOM		&QUALCOMM Inc.					&7739.4		&					&UST		&UST Inc.									&8433.1		\\
CRM			&Salesforce Com Inc. 			&7691.9		&					&GIS			&General Mills							&8243.3		\\
WFR			&MEMC Electronic Materials	&7392.8		&					&KMB		&Kimberly-Clark						&8069.5		\\
\cline{1-3}\cline{5-7}
\\
\multicolumn{3}{l}{Utilities (U)} &~& \multicolumn{3}{l}{Telecommunications Services (TS)}   \\
\cline{1-3}\cline{5-7}
Symbol	&Company							&AMC~ 		&		&Symbol	&Company								&AMC~		\\
\cline{1-3}\cline{5-7}
ETR		&Entergy Corp.						&12798.7	&	 				&T			&AT$\&$T Inc.							&6336.2		\\
EXC		&Exelon Corp.						&9738.8		&					&VZ			&Verizon Communications		&5732.5		\\
CEG		&Constellation Energy  Group &9061.5	&					&EQ			&Embarq Corporation				&5318.7		\\
FE			&FirstEnergy Corp.				&8689.4		&					&AMT		&American Tower Corp.			&5195.6		\\
FPL		&FPL Group							&7742.8		&					&CTL &	Century Telephone		&4333.8		\\
SRE		&Sempra Energy					&6940.6		&					&S			&Sprint Nextel Corp.				&2533.7		\\
STR		&Questar Corp.					&6520.4		&					&Q			&Qwest Communications  Int 	&2201.3		\\
TEG		&Integrys Energy Group Inc. &5978.4		&					&WIN		&Windstream Corporation		&2089.1		\\
EIX		&Edison Int'l							&5877.5		&					&FTR		&Frontier Communications		&1580.9		\\
AYE		&Allegheny Energy				&5864.9		&					&				&												&					\\
\hline
\hline
\end{tabular}
\end{footnotesize}
\end{center}
\label{table4}
\end{table*}

\section{Error estimation}
\label{appB}
Suppose we measured or numerically simulated a set of $M$ data points
$y(\tau_m)$ at positions $\tau_m, \ m=1,\ldots,M$. We want to describe
the data with a function $f(\tau)$ by fitting its $M_P$ parameters.
To assess the quality of the fit, the normalized
$\chi^2$~\cite{Bevington2003}
\begin{equation}
\chi^2 \ = \frac{1}{M-M_P} \sum_{m=1}^{M}\bigl(f(\tau_m)-y(\tau_m)\bigr)^2 \ ,
\label{eq4}
\end{equation}
is used. Here, $M-M_P$ is referred to as the number of degrees of
freedom. In our case, we have $M=1000$, $M_P=3$ for the fitting of trade sign correlators in stock pairs, and $M=34$, $M_P=3$ for the fitting of average trade sign correlators of individual stocks.

\end{document}